\documentclass[3p,times,procedia]{elsarticle}
\usepackage{nupha_ecrc}

\volume{00}

\firstpage{1}

\journalname{Nuclear Physics A}

\runauth{H. Petersen}

\jid{nupha}

\jnltitlelogo{Nuclear Physics A}

\usepackage{amssymb}

\usepackage[figuresright]{rotating}

\begin{document}

\begin{frontmatter}



\dochead{XXVIth International Conference on Ultrarelativistic Nucleus-Nucleus Collisions\\ (Quark Matter 2017)}

\title{Beam energy scan theory: Status and open questions}


\author[label1,label2,label3]{H. Petersen}

\address[label1]{Frankfurt Institute for Advanced Studies, Ruth-Moufang-Stra{\ss}e 1, 60438 Frankfurt am Main, Germany}
\address[label2]{Institut f\"ur Theoretische Physik, Goethe-Universit\"at, Max-von-Laue-Stra{\ss}e 1, 60438 Frankfurt am Main, Germany}
\address[label3]{GSI Helmholtzzentrum f\"ur Schwerionenforschung GmbH, Planckstra{\ss}e 1, 64291 Darmstadt, Germany}

\begin{abstract}
The goal of heavy ion reactions at low beam energies is to explore the QCD phase diagram at high net baryon chemical potential. To relate experimental observations with a first order phase transition or a critical endpoint, dynamical approaches for the theoretical description have to be developed. In this summary of the corresponding plenary talk, the status of the dynamical modeling including the most recent advances is presented. The remaining challenges are highlighted and promising experimental measurements are pointed out. 
\end{abstract}


\end{frontmatter}


\section{Introduction}
\label{intro}
The main goal of the beam energy scan program at the Relativistic Heavy Ion Collider (RHIC) is to find answers to long-standing questions about the phase diagram of strongly interacting matter: 

\begin{itemize} 
\item 
What is the temperature and density? What are the relevant degrees of freedom? 
\item
Is there a first order phase transition between the quark-gluon plasma and the hadron gas and a critical endpoint? 
\item 
What are the transport properties of QCD matter, e.g. $\eta/s(T,\mu_B)$ and $\zeta/s(T,\mu_B)$? 
\end{itemize} 

The experimental programs at RHIC as well as NA61 at SPS and future facilities such as FAIR, NICA and JPARC-HI provide a unique chance to learn something about the QCD thermodynamics that is not (yet) accessible by lattice techniques. To establish definitive links between observables and structures in the phase diagram requires detailed dynamical modeling. Due to the recently obtained and the forthcoming wealth of high precision data in the intermediate beam energy range, there are a lot of theoretical advances that are reflected in this contribution. The focus will be on the beam energy scan program at RHIC. 

In general, there are two regimes with well-established approaches for the dynamical description of heavy ion reactions. At high beam energies ($\sqrt{s_{\rm NN}}>39$ GeV) hybrid approaches based on viscous hydrodynamics and hadronic transport are accepted as the 'standard model' (for reviews see \cite{Hirano:2012kj,Petersen:2014yqa}). The main challenges are to describe the initial non-equilibrium stages and further refine the approaches to perform multi-parameter analyis with Bayesian techniques \cite{Bernhard:2016tnd}. At very low beam energies ($\sqrt{s_{\rm NN}}<3$ GeV) hadronic transport approaches including resonance dynamics and nuclear potentials are very successful in the description of experimental data. The description of the high baryon density stage and multi-particle interactions is still challenging for transport approaches. 

The intermediate energy regime is the one that is of interest here. Due to the time and page limit, only dynamical approaches and recent developments are mentioned. Since physics is smooth, the question is how to interpolate between hybrid and hadronic transport approaches, namely is it possible to surround hydrodynamics with a transport corona or to create transport approaches with 'bubbles' of hydrodynamics? The major challenge in this area is to properly model the phase transition and critical point including non-equilibrium effects. 

\section{Highlights from beam energy scan I}
\label{bes1}
Let us start by recapulating the 5 most interesting results that have been obtained in the beam energy scan stage I program at RHIC. First, the anisotropic collective flow, namely elliptic and triangular flow has to be mentioned \cite{Milosevic:2016kpl,Adamczyk:2016exq}. The surprisingly constant $v_2(p_T)$ can be understand by the interplay between hydrodynamic and hadronic transport phase in hybrid approaches \cite{Auvinen:2013sba}, whereas the triangular flow is more sensitive to the viscosity and might be the better signal of the disappearance of the quark-gluon plasma at low enough beam energies \cite{Karpenko:2016red}. Secondly, the sign change of the slope of the directed flow of protons at midrapidity as a function of beam energy \cite{Adamczyk:2014ipa} has been suggested as a promising signal for the first order phase transition \cite{Brachmann:1999xt,Csernai:1999nf}, but current dynamical models cannot yet achieve a quantitative understanding of the effect \cite{talk_singha_int16_3}.  

The $R_O/R_S$ ratio of HBT radii is supposed to reflect the lifetime of the system, that is significantly prolonged in the case of a first order phase transition. The experimental data from the STAR \cite{Adamczyk:2014mxp} and the NA49 collaboration \cite{Alt:2007uj} both indicate a slight peak in the intermediate beam energy range. Such a peak can be reproduced in hybrid approaches, when a first order phase transition is employed \cite{Li:2008qm}. On the other hand, more recent calculations show that it is difficult to achieve a quantitative description \cite{Batyuk:2017smw}. Furthermore, HBT correlations might provide information on cluster formation. 

The higher moments of the net proton distributions as a function of the beam energy are expected to provide insights on the critical dynamics due to their scaling with higher powers of the correlation length. The correlation length increases when the system passes through the region around the critical point. Recently, the existing results by the STAR collaboration \cite{Luo:2015ewa} have been extended to lower beam energies by the measurements of the HADES collaboration in the recent Au+Au run at $E_{\rm lab}=1.23 A$ GeV (see contributions by M. Lorenz and R. Holzmann). The idea to draw exclusion plots on the phase diagram based on extraced 2-,3- and 4- particle correlations \cite{Bzdak:2016sxg} is intriguing, but at this point needs to await the finalized experimental data and more work on the theory side.  

Last, the very recent new measurement of global polarization of $\Lambda$ hyperons needs to be mentioned. At high beam energies the results are consistent with zero, whereas at lower beam energies a finite polarization with respect to the event plane has been observed \cite{STAR:2017ckg}. On the one hand, this measurement provides constraints on the dynamics of the system, since it implies finite global angular momentum. In addition, by quantifying the difference between the $\bar{\Lambda}$ and $\Lambda$ the order of magnitude of the magnetic field at freeze-out can be constrained, which is of great interest for further investigations of the chiral magnetic effect. The summary of recent developments for chiral anomalies is presented in the contribution by P. Sorensen. 

\section{Initial conditions}
\label{init_cond}

The initial conditions for a hydrodynamic evolution are distributions in coordinate space of the energy, momentum and net baryon density. In principle also the distribution of other conserved quantities, e.g. electric charge or net strangeness is required, if the corresponding currents are conserved during the hydrodynamic evolution. All these distributions fluctuate from event to event. One important question at lower beam energies is, if local equilibrium is established in a large part of the system and at which moment in time. Previous studies assume the geometrical passage time of the two nuclei as the lower limit for the equilibration time. 

\begin{figure}
\centering
\includegraphics[width=0.4\textwidth]{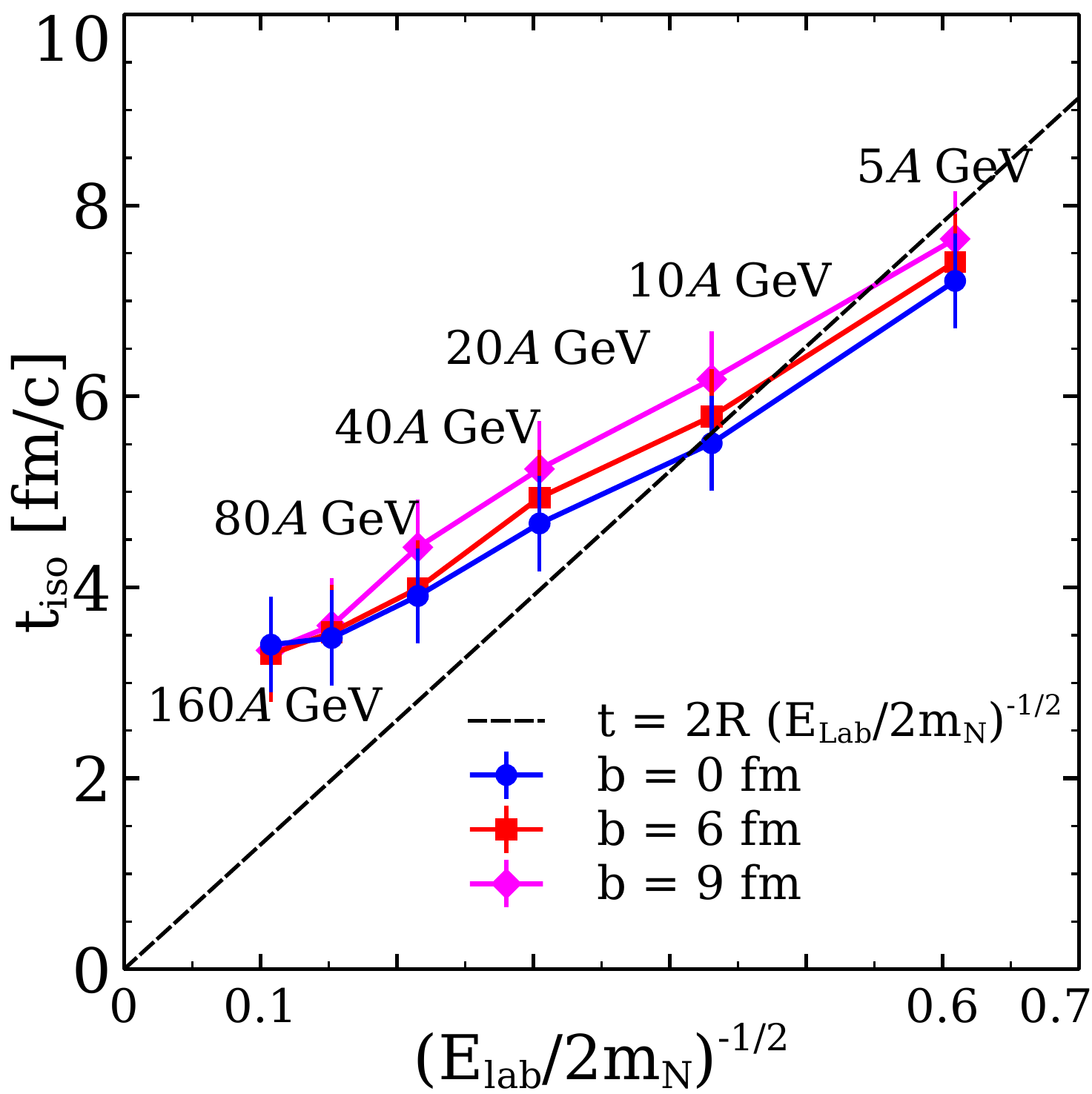}
\caption{Local equilibrium determined in a coarse-grained UrQMD transport calculation with respect to the time, when the two nuclei have passed through each other (taken from  \cite{Oliinychenko:2015lva}). 
\label{fig_local_equilibrium}}
\end{figure}

A recent quantitative analysis of local equilibration in heavy ion collisions at beam energies ranging from $E_{\rm lab}=2-160A$ GeV is shown in Fig. \ref{fig_local_equilibrium}. Within a coarse-grained hadronic transport calculation the isotropization time has been extracted and is plotted as a function of the above mentioned geometrical passing time. At lower beam energies, isotropization is reached faster than the geometrical criterion, whereas at higher beam energies the time of the non-equilibrium evolution is larger. The question, how much non-equilibrium evolution is necessary at intermediate beam energies, still remains to be addressed also in transport approaches including partonic degrees of freedom.  

\begin{figure}
\centering
\includegraphics[width=0.45\textwidth]{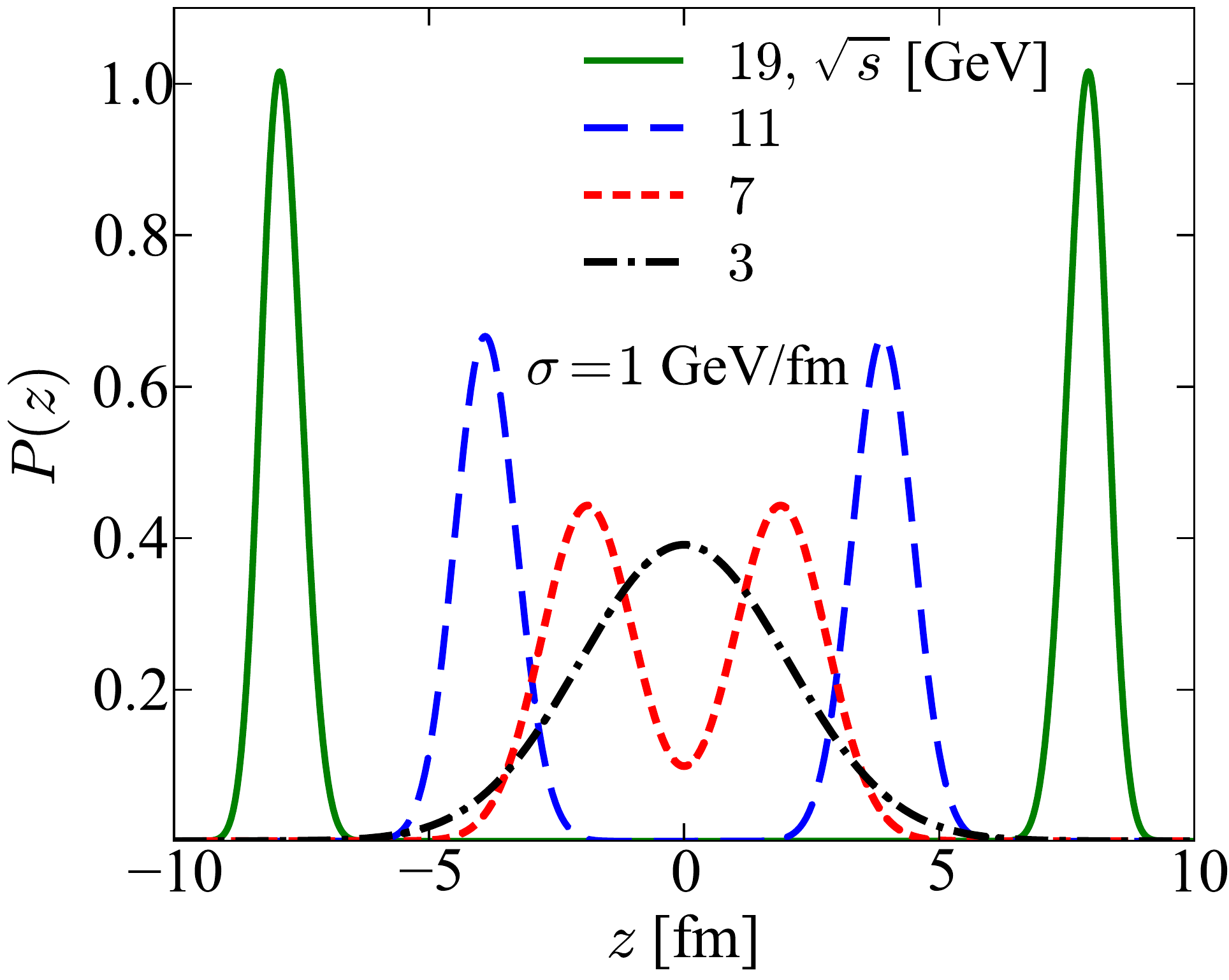}
\includegraphics[width=0.49\textwidth]{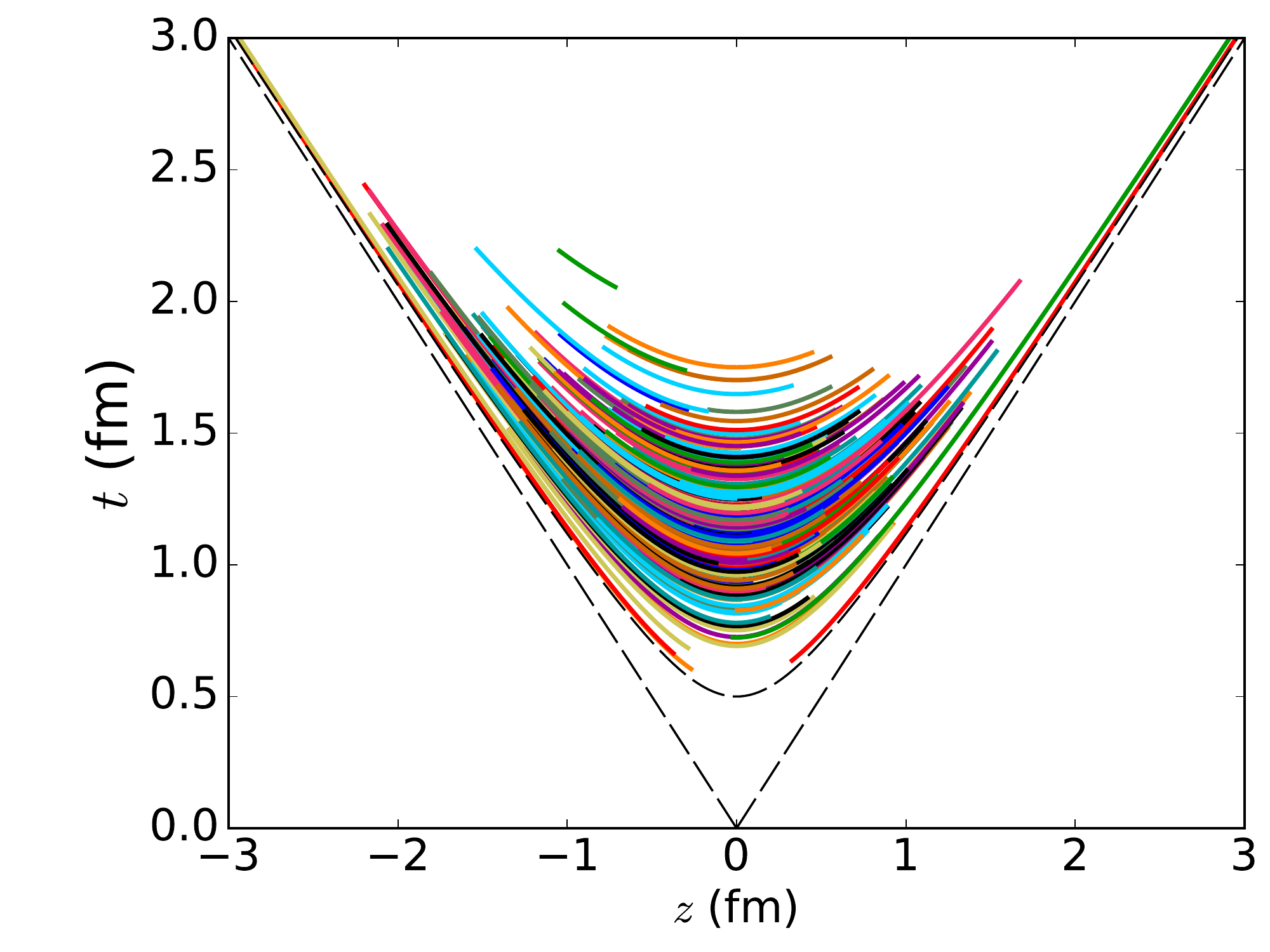}
\caption{Left: The $z$ distribution of stopped nucleons at mid-rapidity for $\sigma=1$ GeV/fm (wounded nucleon model) and various c.m. energies, $\sqrt{s} = 3, 7, 11$ and  $19$  GeV/c (taken from  \cite{Bialas:2016epd}). Right: Distribution of strings in the $t-z$-plane from the LEXUS approach (see talk by C. Shen). 
\label{fig_initial_strings}}
\end{figure}

Another open question is how the energy deposition happens in space-time. Strings are responsible for a large amount of the particle production at higher energies. In Fig. \ref{fig_initial_strings} (left) the distribution of stopped nucleons at mid-rapdity along the beam direction is shown. Within a toy model based on the wounded nucleon picture, a constant deceleration is assumed \cite{Bialas:2016epd}. In contrast to the instant stopping that is implemented in most hadron-string transport approaches, this leads to very different distributions of the initial baryon density. Fig. \ref{fig_initial_strings} (right) displays a new way to construct initial conditions dynamically that has been presented by C. Shen. The Glauber approach is enhanced by the Lexus string model to three dimensions and the energy of the strings is fed into the hydrodynamic evolution by source terms over a finite period of time. 

In general, there are three straight forward options that have to be explored further in the future to achieve a more realistic description of the initial state: 
\begin{enumerate}
\item
Extend Color Glass Condensate approaches to three dimensions and clarify down to which energy the saturation assumptions hold
\item
Employ hadron-string approaches to test, if the string model can reproduce the stopping mechanism 
\item
Investigate if the source terms in 3-fluid-hydrodynamics \cite{Batyuk:2016qmb} with two baryon-rich projectile and target fluids that create the fireball fluid can provide an understanding of the baryon deposition
\end{enumerate} 

All of these possibilities need exploration and confrontation with experimental data. The net baryon density distribution along rapidity has been measured at all beam energies and additional information on the rotation of the system is available from, for example, HBT measurements that provide insights on the source tilt angle \cite{Mount:2010ey}.

\section{Equation of state and transport coefficients}
\label{eos_trans}

\begin{figure}
\centering
\includegraphics[width=0.4\textwidth, angle=270]{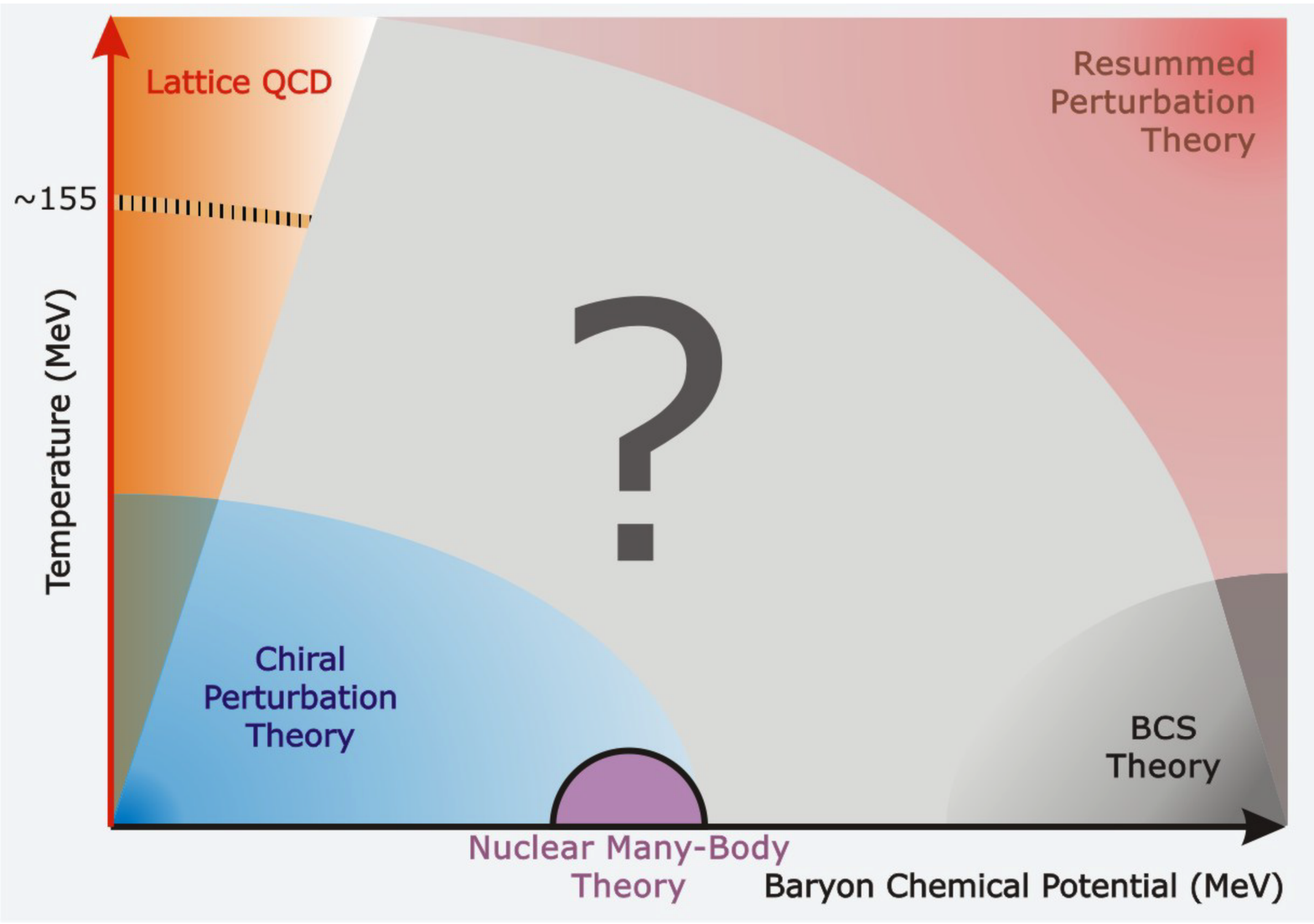}
\caption{Schematic phase diagram displaying the regions covered by first-principle QCD calculations \cite{Steidl_Ffm}. 
\label{fig_phase_diagram}}
\end{figure}

As an external input to the hydrodynamic evolution for the hot and dense stage of the reaction the equation of state and the transport coefficients as a function of temperature and baryon chemical potential have to be known. Fig. \ref{fig_phase_diagram} shows a sketch of the QCD phase diagram indicating the different regimes in which first-principle calculations are available. Lattice QCD calculations predict a cross-over transition up to $\mu_B/T<2.2$. There are independent constraints based on the nuclear ground state properties and in the future possibly from neutron star mergers that can reach temperatures up to $T \sim 50$ MeV. Functional methods, such as Dyson-Schwinger or functional renormalization group calculations, are useful to extrapolate the equation of state into the whole phase diagram. Within such calculations the consistent values for transport coefficients are calculable as well. 

\begin{figure}
\centering
\includegraphics[width=0.52\textwidth]{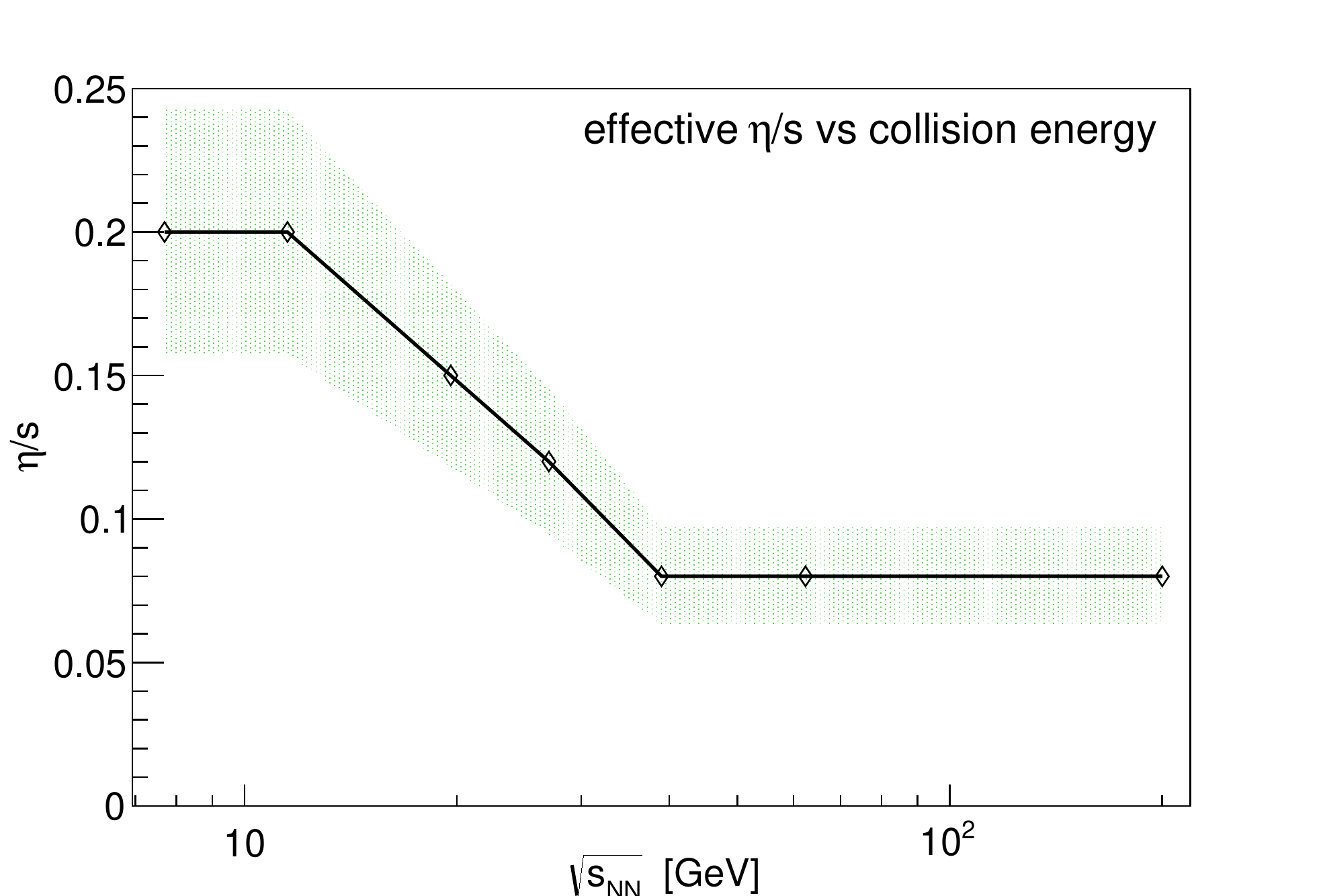}
\includegraphics[width=0.47\textwidth]{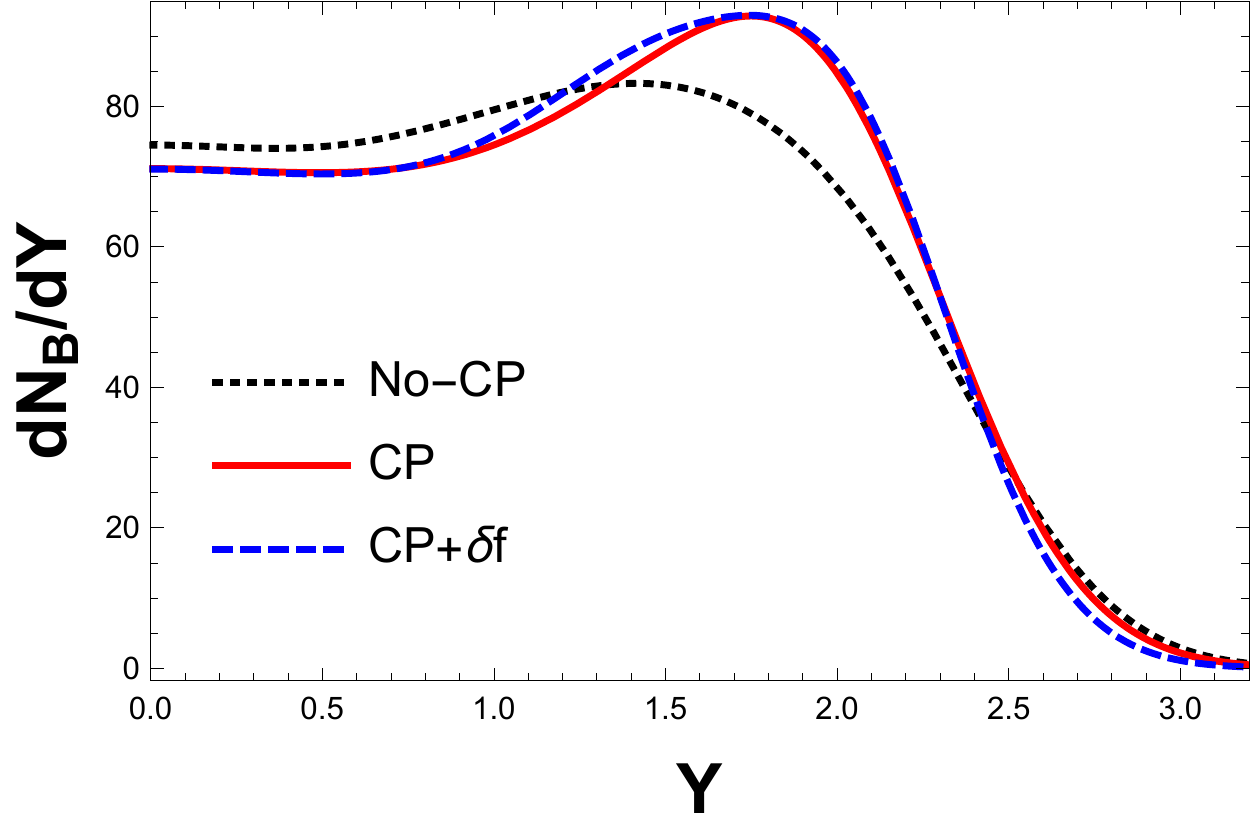}
\caption{Left: Effective ratio of shear viscosity over entropy extracted from a viscous hybrid approach in comparison to bulk observables (taken from \cite{Karpenko:2015xea}). Right:  Net-baryon multiplicity per unit momentum rapidity, $dN_{\rm B}/dY$, as
a function of $Y$. The dashed black, solid red and dashed blue curves, respectively,
correspond to results for hydrodynamical evolution with (CP) without (no-CP) a
critical point, as well as including the non-equilibrium $\delta f$ contribution to
the freeze-out (CP+$\delta f$) (taken from \cite{Monnai:2016kud}). 
\label{fig_shear_bulk}}
\end{figure}

As long as there is no reliable input from first-principle QCD calculations, hybrid approaches can be applied to extract values of the transport coefficients in comparison to experimental data. In this spirit, in Fig. \ref{fig_shear_bulk} (left) the effective shear viscosity over entropy has been extracted by comparing the viscous UrQMD hybrid approach to available bulk observables. Even though this multi-parameter analysis was performed 'by eye', the general trend of an increasing shear viscosity at lower beam energies is reproduced within a recent analysis in the Bayesian framework \cite{Auvinen:2016tgi}.

Fig. \ref{fig_shear_bulk} (right) shows the result of a recent viscous hydrodynamic calculation including bulk viscosity \cite{Monnai:2016kud}. The bulk viscosity is assumed to increase close to the critical point. It can be observed that the net baryon distribution is affected by the influence of the bulk viscosity. As expected the same observable is also influenced by a finite baryon diffusion constant as demonstrated by C. Shen. To disentangle the interplay between different transport coefficients and the initial conditions that also affect the net baryon density distribution, more systematic investigations of many observables in the whole energy range are necessary. 

\section{Non-equilibrium propagation of fluctuations}
\label{non_eq}

\begin{figure}[h]
\centering
\includegraphics[width=0.55\textwidth]{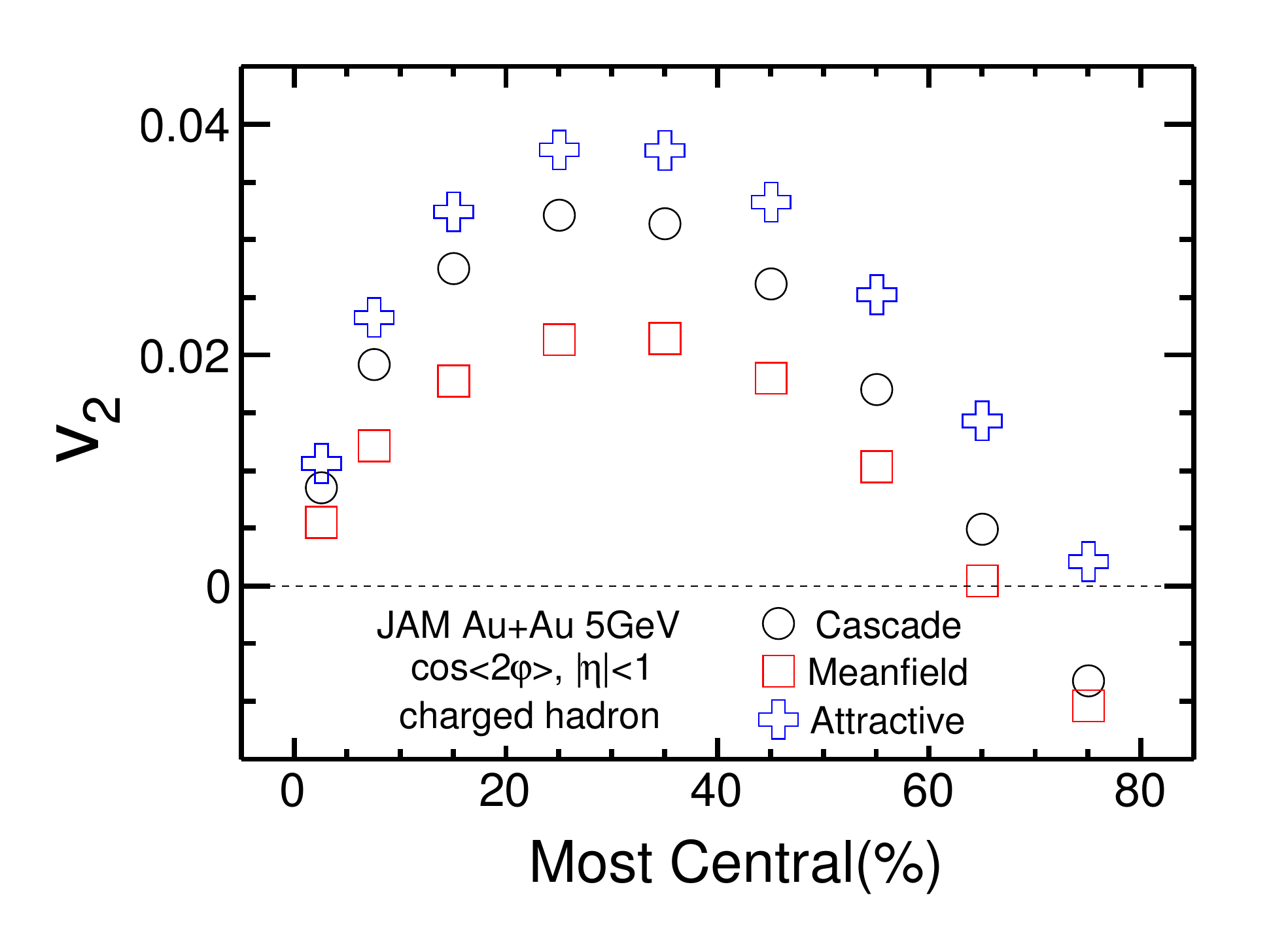}
\includegraphics[width=0.37\textwidth]{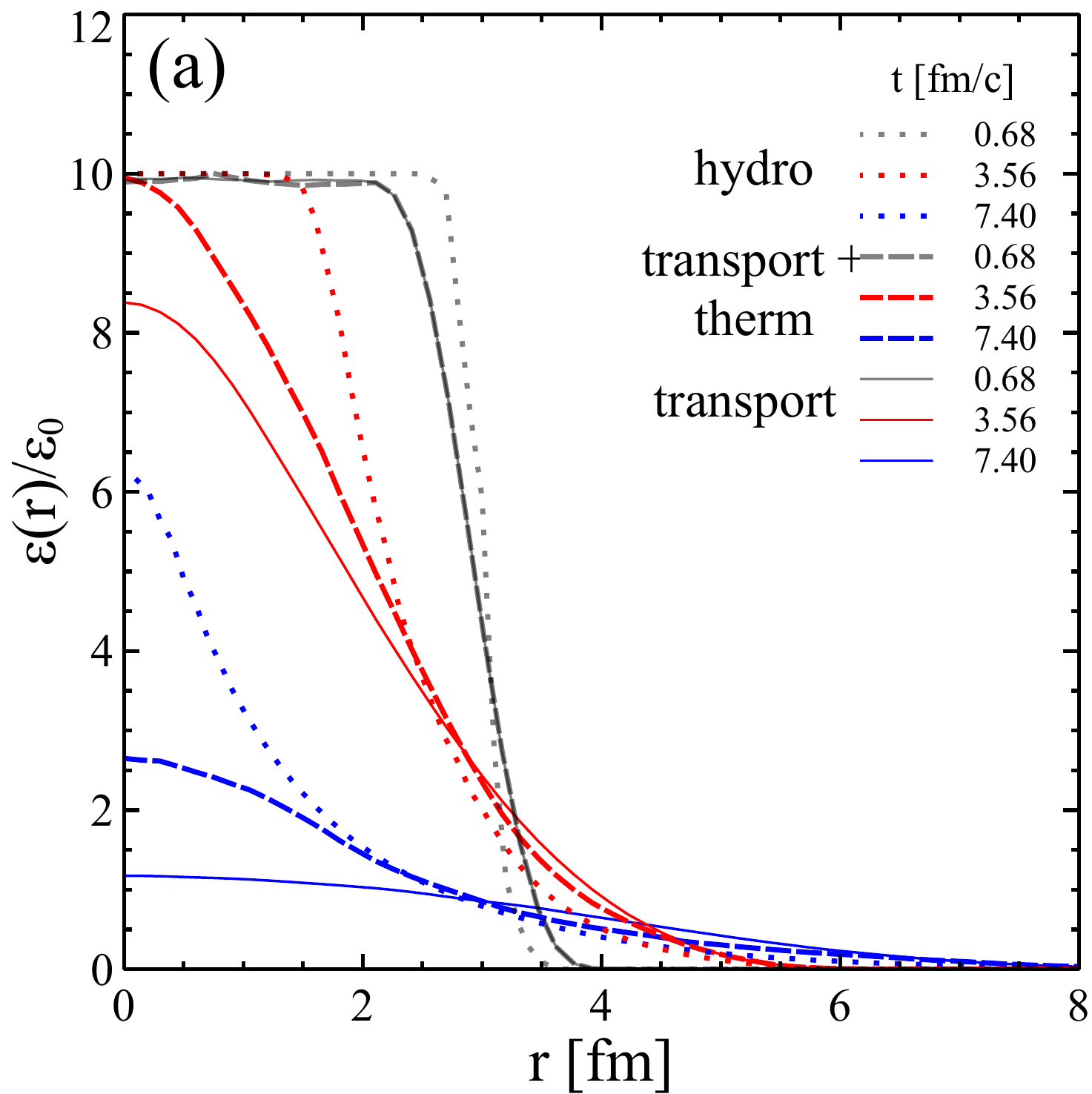}
\caption{Left: The $\eta$ ($|\eta| < 1$) integrated $v_2$ of charged hadrons
 as a function of collision centrality in Au+Au collisions
 at $\sqrt{s_{NN}}=5$ GeV from the standard JAM cascade (circles),
 JAM with mean-field (squares),
 and JAM with attractive orbit (crosses)(taken from \cite{Chen:2017cjw}). Right: Comparison of SMASH pure transport, pure hydro and SMASH with forced thermalization for an expanding sphere (taken from \cite{Oliinychenko:2016vkg}). 
\label{fig_transport}}
\end{figure}

One of the biggest challenges is of course the proper description of non-equilibrium dynamics around a first order phase transition and the critical endpoint. In the previous Sections it was assumed that a hydrodynamic approach is applied to describe the dynamical evolution of heavy ion reactions at intermediate beam energies.

A different option is to start with a microscopic transport approach and include the effects of a phase transition to the quark-gluon plasma. One recent attempt \cite{Chen:2017cjw} to model the changing pressure by different attractive or repulsive interactions within the JAM transport approch is shown in Fig. \ref{fig_transport} (left), where the integrated values of elliptic flow change for the different setups. Fig. \ref{fig_transport} (right) displays the energy density of an expanding sphere as a function of the radius for different timesteps. The calculation has been performed within three different scenarios, one is a pure hadronic transport approach, namely SMASH (Simulating Many Accelerated Strongly-interacting Hadrons) that has recently been developed \cite{Weil:2016zrk}, one is the ideal hydrodynamic evolution with the SHASTA algorithm \cite{Rischke:1995ir} and one is the transport approach with local forced canonical thermalization in high energy density regions \cite{Oliinychenko:2016vkg}. This newly developed technique interpolates between the transport and hydrodynamic evolution and has the advantage that both stages are treated within the transport framework. Also the recently developed viscous anisotropic hydrodynamics (see e.g. \cite{Nopoush:2015yga} and \cite{Bazow:2015zca}) might be useful for lower beam energies.

\begin{figure}
\centering
\includegraphics[width=0.49\textwidth]{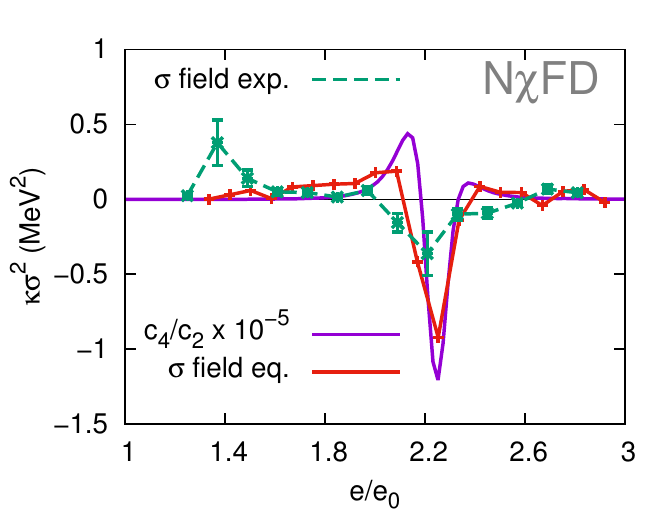}
\includegraphics[width=0.49\textwidth]{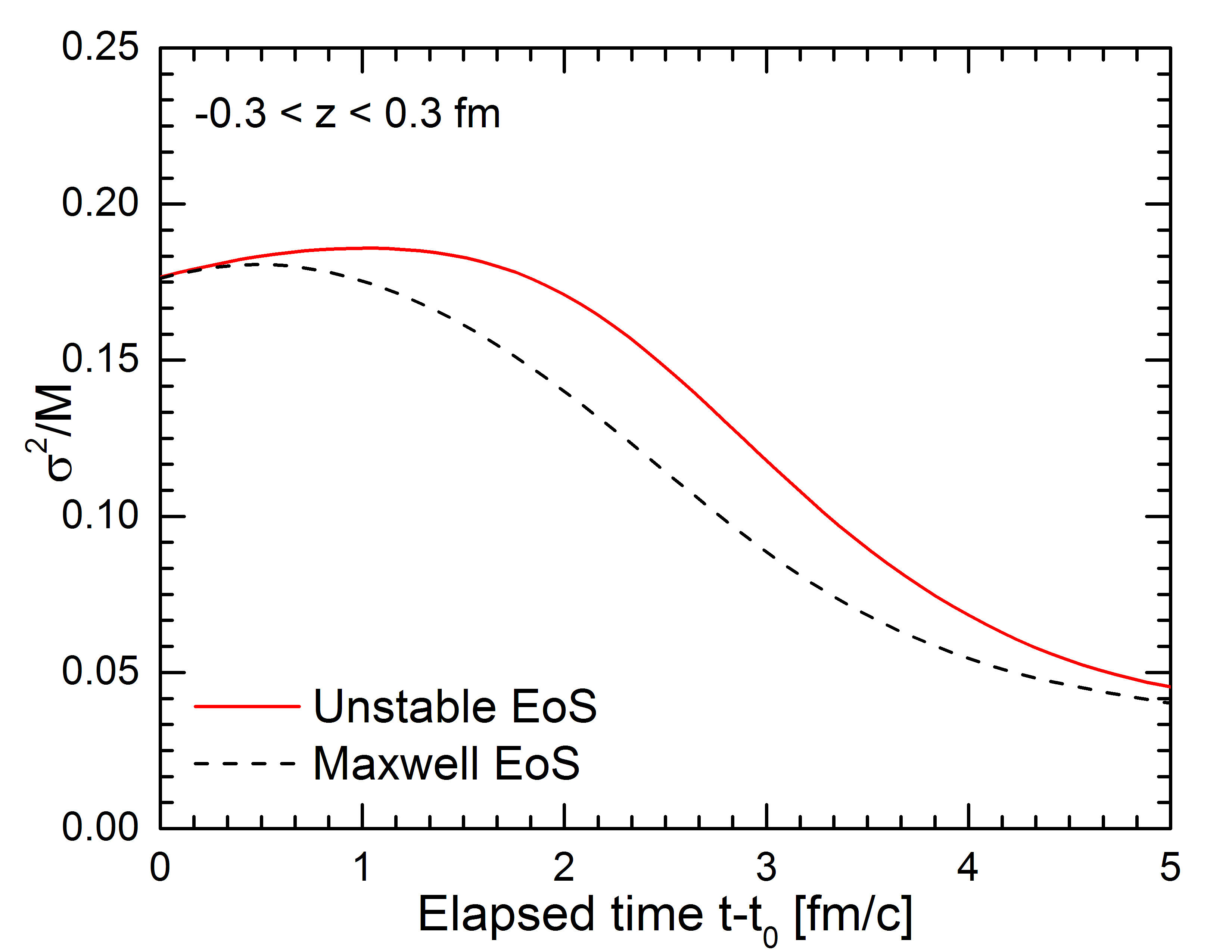}
\caption{Left: Kurtosis as a function of energy density for a trajectory passing close to the critical endpoint in a non-equilibrium chiral fluid dynamics calculation (taken from talk by C. Herold). Right: Variance normalized by the mean of the net baryon distribution from a hydrodynamic calculation including non-euilibrium dynamics at the first order phase transition (taken from \cite{Talk_Steinheimer_INT}). 
\label{fig_non_equilibrium}}
\end{figure}

The qualitative development of fluctuations in the non-equilibrium evolution around a critical endpoint has been studied within chiral fluid dynamics including dissipation and noise (see Fig. \ref{fig_non_equilibrium} (left)). The purple and the red line show calculations of the kurtosis in a box at different values of temperature and baryon chemical potential to verify the agreement of the numerical solution with the thermodynamic expectation. The green curve shows the result of a full non-equilibrium calculation along a trajectory in the phase diagram that corresponds to a beam energy of $\sqrt{s_{\rm NN}}=16.9$ GeV. Overall the main features remain unchanged even though memory effects shift the minimum and maximum of the kurtosis as a function of the energy density. 

There are also multiple attempts to include the noise, the thermal fluctuations, within hydrodynamic calculations directly. Analytic insights on how to treat the different contributions within a hydrodynamic framework have been presented in the parallel talks by M. Stephanov and Y. Akamatsu. M. Nahrgang showed the first steps towards a numerical implementation of 3+1 dimensional hydrodynamics with noise. A net baryon diffusion equation with stochastic noise hase been solved and shown to reproduce the equilibrium expectations. Also non-trivial structures of the kurtosis are dynamically generated and again the minimum is shifted due to memory effects. It is important to note, that such numerical implementations need to be systematically tested to ensure the validity of the results. 

Since the dynamics around the critical point is rather subtle and the higher moments of the distributions are difficult to measure and interpret (see Section \ref{bes1} above), it might be more promising to explore features of the non-equilibrium passage through a first order phase transition. In \cite{Steinheimer:2012gc} the usual van-der-Waals construction for the equation of state has been replaced by the full dependence of the pressure on the energy density allowing the system to develop spinodal decomposition. The resulting density clumps can be observed in an enhancement of the second moment of the net baryon distribution, namely the variance. 

In general, the non-equilibrium dynamics around a critical point or first order transition shows interesting features in moments of the net baryon distribution, but the question, if they survive to be observable still remains. Sticking to full dynamical approaches the Cooper-Frye transition poses challenges to the proper propagation of fluctuations. The enforcement of conservation laws will affect higher moments of conserved quantities and is usually not employed in sampling procedures. Also, the particle sampling on the hypersurface introduces Poisson fluctuations on top of the existing fluctuation signal. The hadronic rescattering further diminishes the amount of correlations that are observable in the final state by another 50 \% \cite{Steinheimer:2016cir}. One way to avoid these complications is to define correlation functions that do not require particlization as suggested in \cite{Pratt:2016lol} or to apply universality arguments \cite{Mukherjee:2016kyu}.

\section{Summary and Conclusions}
\label{summary}

To summarize, there have been a lot of recent developments towards a better understanding of the dynamics of heavy ion reactions at intermediate beam energies. To  provide definitive anwers to the questions about structures in the phase diagram and properties of QCD matter require a detailed understanding of the initial conditions and baryon stopping, the interplay of hadron transport and fluid dynamics, improved input on the equation of state and transport coefficients and further progress on the propagation of fluctuations also through the Cooper-Frye transition and the hadronic stage. Ideally, many observables are addressed within the same approach over the whole beam energy range. Qualitative insights are useful, but there is no way around full dynamical modeling of the non-equilibrium phase transition to obtain firm knowledge about the QCD phase diagram. 

\section{Acknowledgements}
The author is thankful for the presentations and intense discussions during the INT 16-3 program, that inspired a large fraction of this contribution. Especially, I would like to thank my co-organizers of the INT program, V. Koch, M. Lisa and P. Sorensen, and all parallel and plenary speakers of Quark Matter 2017 that provided me with figures and contributed content. 
HP acknowledges funding of a Helmholtz Young Investigator Group VH-NG-822 from
the Helmholtz Association and GSI. This work was supported by the Helmholtz International
Center for the Facility for Antiproton and Ion Research (HIC for FAIR) within the
framework of the Landes-Offensive zur Entwicklung Wissenschaftlich-Oekonomischer Exzellenz (LOEWE) program launched by the State of Hesse.





\bibliographystyle{elsarticle-num}
\bibliography{qm17_petersen_refs}







\end{document}